# Title: Biomimetic peptide enriched nonwoven scaffolds promote calcium phosphate mineralisation

Authors names and affiliation:

Robabeh Gharaei[1]*, Giuseppe Tronci[1,3], Parikshit Goswami[2], Robert P. Wynn Davies[3], Jennifer Kirkham[3] and Stephen J. Russell[1]

[1] Clothworkers' Centre for Textile Materials Innovation for Healthcare, University of Leeds, UK

[2] Technical Textiles Research Centre, University of Huddersfield, UK

[3] Division of Oral Biology, School of Dentistry, St James' University Hospital, Leeds, UK

## Abstract

Cell-free translational strategies are needed to accelerate the repair of mineralised tissues, particularly large bone defects, using minimally invasive approaches. Regenerative bone scaffolds should ideally mimic aspects of the tissue's ECM over multiple length scales and enable surgical handling and fixation during implantation *in vivo*. Leveraging the knowledge gained with bioactive self-assembling peptides (SAPs) and SAP-enriched electrospun fibres, we presented a cell free approach for promoting mineralisation via apatite deposition and crystal growth, *in vitro*, of SAP-enriched nonwoven scaffolds. The nonwoven scaffold was made by electrospinning poly(ε-caprolactone) (PCL) in the presence of either peptide $P_{11}$-4 (Ac-QQRFEWEFEQQ-Am) or $P_{11}$-8 (Ac QQRFOWOFEQQ-Am), in light of the polymer's fibre forming capability and its hydrolytic degradability as well as the well-known apatite nucleating capability of SAPs. The 11-residue family of peptides ($P_{11}$-X) has the ability to self-assemble into β-sheet ordered structures at the nano-scale and to generate hydrogels at the macroscopic scale, some of which are capable of promoting biomineralisation due to their apatite-nucleating capability. Both variants of SAP-enriched nonwoven used in this study were proven to be biocompatible with murine fibroblasts and supported nucleation and growth of apatite minerals in simulated body fluid (SBF) *in vitro*. The fibrous nonwoven provided a structurally robust scaffold, with the capability to control SAP release behaviour. Up to 75% of $P_{11}$-4 and 45% of $P_{11}$-8 were retained in the fibres after 7-day incubation in aqueous solution at pH 7.4. The encapsulation of SAP in a nonwoven system with apatite-forming as well as localised and long-term SAP delivery capabilities is appealing as a potential means of achieving cost-effective bone repair therapy for critical size defects.

## Keywords

# 1. Introduction

Bone or tooth loss due to pathologies such as osteoporosis or periodontal disease continue to represent major healthcare challenges.[1,2] Current therapeutic strategies include bone repair or replacement by device implantation, allogeneic transplantation or autologous bone grafts. However, associated clinical challenges, i.e. multiple number of surgeries and device fixation as well as increasing demand due to the aging population is motivating the need for alternative translational bone repair strategies.

Mineralised tissues in nature are biological composites of calcium phosphates minerals and soft collagen matrices.[3] Hard tissues generally consist of a spectrum of collagen type I fibres and a mineral phase of substituted hydroxyapatite (HAP).[4] Artificial bone replacements or scaffolds for regeneration of mineralised tissue have been realised in the form of composite materials that mimic both the collagen matrix and the mineralised phase of the extracellular matrix (ECM).[5] The most frequently used biocomposite structures for bone regeneration comprise a soft biodegradable polymer scaffold, to provide a similar structure and/or biological function as collagen, and an inorganic phase, such as synthetic HAP, due to its biocompatibility, nontoxicity, and osteoinductive properties.[6-8]

An alternative approach involves mimicking the physiological processes of nucleation and growth of apatite crystals in the ECM of skeletal tissues, aiming to achieve the hierarchical organisation and mechanical properties found in bone *in vivo*.[9] Research has therefore been carried out to generate bioactive scaffolds with the ability to promote nucleation and support HAP crystal growth similar to that seen in ECM.[10,11] Apatite forming capability of polymer scaffolds can be evaluated by incubating them in a near-physiologic conditions, e.g. in simulated body fluid (SBF) *in vitro,* which has an ion concentration comparable to that of human blood plasma. Any nucleated crystals can then be characterised.[12,13]

The 11-residue family ($P_{11}$-X) of self-assembling peptides (SAPs) when incubated in near-physiologic conditions, generates three-dimensional hydrogel scaffolds that mimic the organic chemical composition of the ECM of hard tissues and are capable of inducing apatite deposition *in situ*.[14] $P_{11}$-X SAPs consist of hydro-gelating amino acidic sequences that assemble into hydrogen bonded β-sheet tapes (a single-molecule thick) above a critical concentration (C*)[15], with higher order structures (ribbons, fibrils and fibres) being realised if the concentration is further increased.[16] This class of peptides also undergoes self-assembly when subjected to external stimuli such as variation in pH[17] and ionic strength[18], which can be exploited in e.g. stimulus-triggered drug delivery. Among the $P_{11}$ category of peptides, $P_{11}$-4 and $P_{11}$-8 exhibit an overall net charge of -2 and +2, respectively, and have been demonstrated to display low immunogenicity *in vivo* and no cytotoxic effects in human and murine cells.[16,19-21] They have promising properties for biomedical applications and tissue regeneration in glycosaminoglycan-depleted tissues including cartilage, in soft tissues and in bone.[14,20,22-27]

Most previous studies using $P_{11}$-X peptides have focused on the molecular design of self-assembled hydrogels, whereby promising results have been shown with regard to cell growth and hard tissue repair[14,16,20, 21,25,28-30]. Kirkham et al. [14,24] evaluated $P_{11}$-4 in treating early dental caries (decay) wherein the peptide was found to promote nucleation and growth of HAP *in situ* within the lesions.

Moreover, $P_{11}$-4 and $P_{11}$-8 hydrogels were applied *in vivo* to critical size defects in rabbit calvaria and bone regeneration was observed over three months, with mineralised tissue ingrowth observed to be greater in calvaria treated with negatively charged $P_{11}$-4- than positively charged $P_{11}$-8- samples.[26] In another *in vivo* study, it was shown that $P_{11}$-4-based hydrogel accelerated healing in calvarial defects in rats.[31] This was attributed to the ability of $P_{11}$-4 to act as a heterogeneous nucleator in the assembled form through negatively charged domains that attract calcium ions and form the nucleus for HAP deposition.[31]

An important requirement for resorbable bone tissue scaffolds is to maintain a stable structure over time capable of promoting regeneration via progressive apatite crystal deposition and crystal growth.[32] Despite the biofunctionality of self-assembled $P_{11}$ peptide-based gels, they frequently suffer from poor strength and lack of dimensional confinement *in situ* when used in regenerative medicine applications. This can make their handling and surgical fixation challenging in large load-bearing tissue defects. The incorporation of SAPs within PCL electrospun fibres can overcome these challenges while retaining the scaffolds' bioactivity[33-35] and we have previously demonstrated that polymer-assisted peptide electrospinning successfully enables the formation of molecularly-assembled peptide nanofibres in an electrospun nonwoven.[36-38] $P_{11}$-8-supplemented electrospun PCL webs were also found to be biocompatible with mouse fibroblasts.[37,38]

In this work, we investigated the apatite induction capability of both $P_{11}$-4- and $P_{11}$-8-supplemented PCL nonwoven scaffolds and characterised the peptide release behaviour, fibre morphology and wettability, as well as the cytotoxicity of both variants. The nonwoven scaffolds were also studied in SBF to explore their capacity to nucleate and grow HAP crystals. Furthermore, we quantified the release behaviour of the SAPs component of the nonwoven scaffolds when they were physically dispersed in aqueous media at different pHs, aiming to evaluate the potential retention of peptides within any defect site as this is an important factor to support the bone regeneration process.[32]

## 2. Materials and methods
### 2.1. Preparation of solutions and electrospinning

PCL ($M_n$: 80,000 g.mol$^{-1}$) and HFIP (purity ≥ 99.0%) were purchased from Sigma Aldrich UK. Peptides $P_{11}$-8 (CH$_3$CO-Gln-Gln-Arg-Phe-Orn- Trp-Orn-Phe-Glu-Gln-Gln-NH$_2$) (Mw =1,565, a peptide content of ~ 75%, and a HPLC purity of 96%) and $P_{11}$-4 (CH$_3$CO-Gln-Gln-Arg-Phe-Glu-Trp-Glu-Phe-Glu-Gln-Gln-NH$_2$) (Mw =1,595, a peptide content ~ 94.9% and a HPLC purity of 95.0%) were purchased from CS Bio Co. USA. Electrospinning solutions of 6% (w/w) PCL in HFIP with either 10, 20 or 40 mg mL$^{-1}$ of $P_{11}$-4 were prepared according to the method explained previously.[38] Additionally, a PCL solution supplemented with 40 mg mL$^{-1}$ $P_{11}$-8 were prepared for biomineralisation studies. All of the solutions were spun using a standard single spinneret electrospinning setup and other processing parameters were identical to those previously reported.[38] Sample nomenclature is defined in Table 1.

### 2.2. Scanning Electron Microscopy (SEM & EDX)

Dry electrospun samples were sputter coated with platinum with a thickness of 8 nm and imaged using a field emission gun scanning electron microscope (LEO1530 Gemini). The microscope was also fitted with an energy-dispersive X-ray spectrometer (EDX) of Oxford Instruments AztecEnergy to investigate the chemical composition of the mineralised crystals on the scaffolds after biomineralisation assay.

### 2.3. Surface Wettability and Direct Cytotoxicity assay

The degree of hydrophilicity and biocompatibility of scaffold samples 2-4 were assessed using goniometry (FTA 4000 Microdrop®) and a contact cytotoxicity assay using murine L929 cells respectively. Results were compared with previously published data for samples 1 and 5 obtained using identical methods.[38]

### 2.4. Peptide Release from PCL/peptide Scaffold Samples

Peptide-supplemented scaffold samples (3 replicates) with an original peptide concentration of 40 mg mL$^{-1}$ (sample 4 and 5) along with control sample 1 were incubated in water and titrated to different pH values: 3.5, 7.5 and 10.5 at 25°C. The purpose was to determine the kinetics of peptide disassembly and release, enabling the stability of the peptides in the fibres to be evaluated under simulated biological conditions. The ratio of fibre to buffer solution was 1 mg in 3 mL (calculated based on the critical concentration $C^*$ for self-assembly of the peptides[16]) and the samples were incubated at 120 r min$^{-1}$ in an orbital shaker incubator (ES-20 model from Grant Bio) at 37°C. After 1 hour, 24 hours, 48 hours and 168 hours the samples were washed three times in the same fresh solution of identical pH to remove all remaining peptide residues and were dried in a desiccator. Overall mass loss from the scaffolds was then determined gravimetrically after incubation, based on the dry state.

### 2.5. Circular Dichroism Spectroscopy (CD)

To analyse the release of peptides from fibres into solution, CD analysis of the supernatant fluid was conducted and spectra were recorded using a Chirascan CD spectrometer with 1 mm path-length cuvettes at 22°C. The data were acquired at a step resolution of 1 nm and scan speed of 60 nm min$^{-1}$. A bandwidth of 4.3 nm was used to obtain smoother spectra. Far-UV spectra were recorded in the wavelength range 185 to 260 nm. Each spectrum was the average of two scans and the spectrum for the blank solvent was subtracted. The data then were converted to mean residue ellipticity (deg cm$^2$ dmol$^{-1}$) and fitted with a polynomial equation ($R^2 \geq 0.95$).

### 2.6. Incubation of fibres in simulated body fluid (SBF)

The apatite-forming capability of scaffold samples containing peptide (numbers 4 and 5) was evaluated *in vitro* in SBF based on the international standard method of BS ISO 23317:2014 and compared with the control sample 1. The electrospun PCL/P$_{11}$-4 and PCL/P$_{11}$-8 samples (3 replicates) were cut out (70±10 mm × 70±10 mm) and weighed with 40 mg of peptide available in each sample, calculated based on the concentration and purity of each peptide in the spinning solution. SBF was prepared in accordance with the standard method in ion-exchanged and distilled water analytical grade chemicals listed in SI Table 1.[39]

The pH of the SBF solution was adjusted to 7.4 ± 0.1 at 36.5 ± 0.2 °C. Samples were incubated at 36.5°C in SBF with a ratio of 0.5 mg mL$^{-1}$, as suggested by Poologasundarampillai et al.,[40] at three time points of 1, 2 and 4 weeks, and then gently rinsed with distilled water. Unwashed specimens were also retained to determine the effect of washing. Finally, the samples were dried in a desiccator at room temperature for 48 hours.

### 2.7. X-Ray Diffraction Crystallography (XRD)

XRD was selected for this study as it is a well-established non-destructive technique for the identification of HAP and also to confirm the EDX and SEM analysis. Samples were examined by XRD (PANalytical X'Pert MPD fitted with X'pert Highscore Plus software) to determine the molecular structure of crystalline material formed on scaffolds during incubation in SBF. Samples were placed as flat sheets in the sample holder and data collection was based on the 2θ scan method using Cu Kα radiation. Spectra were collected between 5° and 65° with a step size of 0.05° and accelerating voltage of 45 kV.

### 2.8. Statistical analysis

Statistical analysis was carried out in relation to the differences in the Ca:P ratios after incubation in SBF. A single factor ANOVA test was carried out to compare the results of the three time points (1, 2 and 4 weeks) and a t-test to determine any effect of the washing process on the results of paired data (washed and unwashed).

## 3. Results and discussion

Electrospun scaffolds of PCL with varying concentration of $P_{11}$-4 were successfully produced and their fibre morphology, water contact angle and cytotoxic response were compared with our previously reported PCL/ $P_{11}$-8 scaffolds.[38] Moreover, the peptide release kinetics and apatite-nucleation/crystal growth capability of both $P_{11}$-4- and $P_{11}$-8-supplemented scaffolds were analysed to elucidate their potential to achieve a long-term therapeutic delivery of SAPs to repair critical size bone defects.

The present work investigates the mineralisation capability *in vitro* of the $P_{11}$-4 and $P_{11}$-8-loaded PCL fibres, as a first step to assess their applicability in guided bone regeneration therapy. Building on the bioactivity of the above-mentioned peptides [14,20-22,24-26,31], the materials developed in this study could therefore find applications in dental tissue repair or as a barrier membrane to stimulate the repair of critical size bone defects (CSBDs) (Schematic shown in Figure 1). In the first instance, an advanced caries lesions with the total breakdown of the enamel, known as 'cavity' can be filled with a customised membrane scaffold, benefitting from surgical handling capability, excellent biocompatibility and easy customisation of the material in terms of shape and size at the chair side for different shapes of cavities. The scaffold material then can be fixed using a biodegradable tissue glue e.g. cyanoacrylate glue within the cavity[41] and promote repair by providing a biomimetic scaffold capable of HAP nucleation, and mineralised tissue growth. In the second proposed application, the CSBD is filled with an autograft or a synthetic bone graft that supports the weight strain in case of a weight-bearing defect, and then the defect site can be wrapped with the membrane for guided bone regeneration (GBR) and to direct the growth of new bone. The membrane architecture, porosity and thickness can be customised to act as barrier to prevent soft tissue growth and invasion from outside and the inner side can promote mineralised tissue ingrowth due to SAPs availability at the surface. The membrane can then be fixed using degradable screws for the orthopaedic fixation,[42] and it can also be designed to be degradable within a suitable profile.

### 3.1. Fibre morphology and wettability

Figure 2A-D are micrographs of nonwoven samples 1-4 electrospun with varying concentrations of $P_{11}$-4 (0-40 mg mL$^{-1}$). The addition of $P_{11}$-4 into pure PCL spinning solution results in a bimodal superimposed nanofibrous network similar to that previously observed in the spinning of $P_{11}$-8-supplemented electrospinning PCL solutions.[38]

As observed previously with $P_{11}$-8-supplemented PCL fibres,[38] relatively small contact angle values of 58±1° and 70±1° were observed in peptide-enriched nonwovens when the concentration of peptide increased from 10 to 20 mg mL$^{-1}$, sample 2 and 3 respectively (SI Figure 1A-B). In contrast, a contact angle of 130 ± 0.5° has been observed for 100% PCL electrospun fibre webs,[38] providing supporting evidence of the peptide-induced increase in scaffold wettability in deionised water. Figure 2E illustrates the dynamic contact angles of deionised water on samples 2 and 3 where wetting progressively increased as a result of higher peptide content. At the higher $P_{11}$-4 concentration of 20 mg mL$^{-1}$ (sample 3), the rate of wetting was so rapid that after 3 s complete penetration of the droplet occurred. This was in contrast to the hydrophobic response of the 100% PCL sample, where the contact angle was invariant over the 15 s test period.[38] At 40 mg mL$^{-1}$ peptide concentration (sample 4), the rate of droplet penetration was so rapid that a contact angle could not be determined.

In line with the aforementioned contact angle observations, Figure 2F confirms that the L929 murine fibroblasts proliferated in direct contact with $P_{11}$-4-enriched fibres (sample 4) with no evidence of cytotoxicity. The L929 cells were selected to conform with the ISO standards related to the cytotoxicity evaluation of medical devices (EN DIN ISO standard 10993-5). These findings are comparable to those observed when $P_{11}$ peptides were cultured in DMEM.[38] The quantitative results from the direct cytotoxicity assay (SI Figure 1C) also revealed that the number of living cells in the PCL/$P_{11}$-4 (sample 4) and PCL/$P_{11}$-8 (sample 5) were comparable to that of the PCL negative control sample ($p \geq 0.05$) and the DMEM control. This is to be expected due to the non-cytotoxic nature of PCL and self-assembled peptides during the time of cell culture period. The mean optical density (OD) of the samples was used to calculate the percentage of cell viability based on the equation reported by Park et al[43] (SI Table 2). PCL/$P_{11}$-4 (sample 4) and PCL/$P_{11}$-8 (sample 5) showed at least 84% and 77% cell viability, respectively, compared to the DMEM control, and up to 100% and 92% cell viability compared to the PCL control.

### 3.2. Peptide release kinetics from PCL fibres

Maintainence of a stable structure is an important requirement for bone tissue scaffolds, while newly regenerated tissue is formed and bone mineralisation takes place[31]. In an *in vivo* study by Burke et al., $P_{11}$-4 and $P_{11}$-8 hydrogels injected into full thickness calvarial defects in rabbits supported bone repair which was complete in some cases and incomplete in others.[22] Evidence of full peptide hydrogel diffusion (i.e. complete depletion) was reported within some of the defects by day 7, which may have contributed to the incomplete defect repair. In the current study, the physical incorporation rather than covalent immobilisation of the peptides within a degradable PCL nonwoven has the potential to modulate the rate of peptide diffusion and therefore further support the bone regeneration process *in vivo*.

Moreover, it is paramount to consider the effect of pH on the stability of bone scaffold and peptide release kinetics. The SAPs studied herein are known to be pH-sensitive, such that they self-assemble into fibrils in water depending on the pH and peptide concentration in solution. In principle, all of the transformations from monomeric random coil to self-assembled fibrils are reversible[17,44] and as such, all incubation experiments were carried out in aqueous media at acidic, neutral and basic pHs.

Peptide release from the fibres was determined gravimetrically before and after incubation in water. Figure 3A shows that the overall mass loss of peptide containing fibres increased over a period of 168 hours (7 days). However, the mass of PCL control fibres did not change over the same period of time regardless of the pH of the solutions. Therefore, the overall mass loss in the PCL/peptides fibres is likely to be due to the release of peptides from the fibres into solution.

As expected, the highest mass loss in $P_{11}$-4-supplemented nonwoven scaffolds took place after 168 hours of incubation at pH 10.5 (up to 9.3%), and for PCL/$P_{11}$-8, after 168 hours of incubation at pH 3.5 (up to 14.5%). Note that these are the least favourable pH values for each of the two peptides respectively, i.e. the pH values at which each of the peptides are expected to transform from self-assembled β-sheet form (fibrils and fibres) to monomers (non-Newtonian fluid).[44] Based on the peptide purities and solution concentrations of 40 mg mL$^{-1}$, the initial weights of peptide in the PCL/P$_{11}$-4 and PCL/P$_{11}$-8 scaffolds prior to incubation were 28% (w/w) and 24% (w/w) respectively. Therefore, in the PCL/P$_{11}$-4 fibres, only 33% of the original P$_{11}$-4 content was presumed lost over 168 hours, and for the PCL/P$_{11}$-8 samples it reached 60%, in the least favourable conditions.

The opposite net charge of P$_{11}$-8 (+2) and P$_{11}$-4 (-2) was also found to play a key role on the scaffold's release capability; P$_{11}$-8-supplemented nonwoven samples displayed the highest apparent release of peptide in acidic conditions, whereby the 168 hours cumulative release was decreased at both neutral and basic pH; whilst the opposite trend was observed with P$_{11}$-4-supplemented nonwoven samples. Aforementioned trends in release kinetics putatively reflect the effect of peptide electrostatic charge and respective electrostatic interactions with solution ions at the pHs investigated.

At a pH of 7.5, which is closest to biological conditions (pH 7.4), 75% of the P$_{11}$-4 and 45% of the P$_{11}$-8 apparently remained within the electrospun fibres after 168 hours. This is noteworthy given how quickly the same peptides were reported to diffuse from a treatment site when delivered in the form of a hydrogel.[22]

The fibres were also analysed microscopically to monitor morphological changes, e.g. fibre size, which can be expected to happen after diffusion of the SAPs to the solutions. SI Figure 2 shows SEM micrographs of 100% PCL, P$_{11}$-4 and P$_{11}$-8-supplemented nonwovens (sample 1, 4 and 5 respectively), incubated for up to 168 hours at the lowest and highest levels of pH of 3.5 and 10.5 respectively. No marked differences in general nanofibrous architecture were observed for the P$_{11}$-4 or P$_{11}$-8-supplemented nonwovens (and control PCL) after 168 hours of incubation other than a flattened fibre morphology and a minor swelling, most probably due to water absorption. Consequently, the observed difference in the level of retained P$_{11}$-8 and P$_{11}$-4 peptides is likely to reflect the effect of peptide-induced electrostatic interactions with solution ions, as well as the different fibre dimensions in the two fibrous scaffolds, which will influence the length of the internal diffusion pathways. A larger proportion of the fibres in the P$_{11}$-8 sample were below a diameter of 100 nm promoting rapid peptide diffusion, compared to samples containing P$_{11}$-4 peptide.[38]

To confirm release of peptides into the solution, the supernatant fluid of 100% PCL, P$_{11}$-4 and P$_{11}$-8-supplemented scaffolds which were incubated at the lowest and highest values of pH; 3.5 and 10.5 respectively, for 24 hours and 168 hours, were analysed. CD analysis was then conducted and the results are shown in Figure 3B-C, with the spectra of the associated blank solution subtracted. All the CD spectra for the P$_{11}$-4 and P$_{11}$-8 supernatants exhibited a negative minimum band at 195-200 nm and a slightly positive band at around 220 nm, which is consistent with disassembly of the peptides into a random coil conformation in solution.[16,45] This explanation is supported by the lack of a negative band at 218 nm, and a strong positive maximum at lower wavelengths (198 nm), which would be characteristic of β-sheet conformation.[16,46] Comparing these results with those of the PCL fibre supernatant, which exhibited no peaks in the far-UV spectral region (190-250 nm), it can be confirmed that peptides are disassembled into solution over time. Moreover, the peak intensities in the CD spectra support the results of the mass loss experiments, in which the highest mass losses obtained for the $P_{11}$-4 and $P_{11}$-8-supplemented nonwovens after 168 hours of degradation were observed at pH 10.5 and pH 3.5 respectively.

### 3.3. Apatite-nucleation ability of fibres following incubation in SBF

Representative SEM micrographs of scaffold samples incubated in SBF for one, two and four weeks' duration are given in Figure 4. After one week's incubation, no apatite crystals could be detected on $P_{11}$-8-supplemented scaffolds (sample 5) (Figure 4J) and only a small amount of crystals was observed in $P_{11}$-4-supplemented scaffold (sample 4) (Figure 4D,G). However, increasing the incubation times in SBF to 2 weeks and 4 weeks (Figure 4E, F, H, I, K, L) led to substantial crystal nucleation and growth, both on the surface of, and within the pore structure of the peptide loaded scaffolds.

No evidence of apatite crystal formation was observed on the PCL control samples containing no peptides, even after the longest incubation time of four weeks (Figure 4C). Apatite crystals formed on the peptide-loaded fibre surfaces and within the proximal pore network exhibited a globular, cauliflower-like shape, resembling clusters of HAP crystals.[40] It may be assumed that a primary layer of calcium phosphate is formed over the surface of the fibres in the scaffold, over which further growth of spheroidal clusters proceeds.

Low magnification SEM micrographs (1000-3000 x) of $P_{11}$-4-supplemented samples incubated in SBF for one, two and four weeks, revealed a distribution of mineralised crystals all over the fibrous structure, Figure 4G-I. The frequency of crystal formation within the fibrous structure increased with incubation time, and the scaffold's` fibrous structure provided a supporting matrix for apatite crystals, promoting their attachment and growth.

#### 3.3.1. Atomic ratio of Ca:P via energy-dispersive X-ray (EDX) analysis

The increase in amount of crystal deposits on the SAP-supplemented scaffolds with incubation time was also confirmed by elemental analysis (EDX). SI Table 3 and SI Figure 3 report representative data and the EDX pattern for sample 4 incubated in SBF for 2 weeks and a calcium to phosphorus ratio of 1.72 is revealed, which is broadly consistent with apatite formation at the surface.[47]

All of the incubated PCL and SAPs-supplemented nonwovens showed peaks corresponding to C and O, together with minor peaks for Na, Cl, K and Mg, attributable to ion precipitation. However, samples comprising PCL/peptides also showed characteristic peaks for phosphorus (2.01 eV) and calcium (3.69 eV) suggesting the presence of apatite. The mean Ca:P molar ratios for all samples before and after washing are summarised in Table 2 based on EDX analysis of random positions on individual mineral particles (n=3 per sample).

In the PCL control sample, no mineral deposition was observed, as confirmed by visual observations in the SEM study. The Ca:P molar ratios in the $P_{11}$-4-supplemented scaffolds (sample 4) from week 1 of incubation was close to 1.67, which is indicative of HAP.[40,47] The Ca:P molar ratio for these samples were comparable at different time points and no significant difference was observed (P >0.05). These data also suggest that the washing process did not markedly affect the Ca-P mineral deposition on $P_{11}$-4-supplemented samples (P>0.05).

In the $P_{11}$-8-supplemented scaffolds (sample 5), the Ca:P ratio of SBF-incubated samples was found to be directly related to the incubation time in SBF. A Ca:P ratio of 1.65 was measured in retrieved samples only after four weeks, whilst decreased and highly variable values of Ca:P ratio were observed at earlier time points, suggesting a lower and uncontrollable mineralisation capability of $P_{11}$-8- with respect to $P_{11}$-4-enriched fibres. Furthermore, the average Ca:P ratios in samples composed of $P_{11}$-8 appeared to be decreased in retrieved samples following washing in water at all the time points, whereby statistical analysis of washed and unwashed results at week 4 showed significant difference in Ca:P ratios (P < 0.05). This suggests the deposition and precipitation of calcium phosphate species on the fibre surface in addition to the mineralisation of calcium phosphate crystals.

The difference in the above results may be attributable to differences in the morphology of the $P_{11}$-4-supplemented scaffolds compared to those of $P_{11}$-8.In the latter case, a biphasic structure exists, with a larger number of peptide-enriched nanofibers that bridge the pores between larger submicron fibres.[38] Owing to their nanofiber dimensions and large surface area, a shorter diffusion pathway for the peptide component at the fibre surfaces may be expected, and therefore a more rapid dissolution. By contrast, the $P_{11}$-4-supplemented scaffolds showed greater control over mineralisation. This could be due to either the effect of the negatively-charged surface, promoting a template for stabilisation of calcium ions; or the presence of fewer nanofibers, suggesting that the peptide is mainly incorporated within larger diameter fibres, Figure 2. Any minerals associated with the nucleating site of the peptide will therefore also be attached to a more stable substructure, and less likely to be removed by washing. Moreover, the superficial washing out of peptide at the surface of these fibres will be replenished and this will allow for further exposure of active groups towards the fibres' surface, thereby creating an area of high crystal growth support.

The spheroidal morphology of the calcium phosphate crystals on the SAPs enriched fibres may arise from the nature of the nucleating site in the peptides and their affinity for mineral ions. As both peptides have a net charge (-2 or +2), it is suggested that they drive mineralisation through either cation or anions ion attraction (calcium or phosphate) thereby providing a site for crystal growth.[48] Moreover, the molecular organisational structure of the peptides into fibrils and presentation of charge domains at the fibrillar surface will influence the capability of assembled peptides to nucleate and support crystal growth *in vitro*.

### 3.3.2. X-ray diffraction analysis

X-ray diffraction patterns of electrospun 100% PCL, $P_{11}$-4 and $P_{11}$-8-supplemented samples after two and four weeks are presented in Figure 5. The data are normalised and two distinct diffraction peaks were observed in all samples at $2\theta = 21.5°$ and $2\theta = 23.8°$, which were indexed to the crystalline structure of PCL.[49,50] In the PCL control (sample 1) patterns (SI Figure 4), no additional peaks were observed after incubation in SBF. By contrast, incubation of PCL/peptide samples in SBF (Figure 5) all led to additional X-ray diffraction peaks.

The peak at $2\theta = 31.8°$ and an isolated peak at $2\theta = 46.6°$, in $P_{11}$-4 and $P_{11}$-8-supplemented nonwovens after incubation is indicative of the presence of HAP.[51] The patterns relating to HAP in the PCL/peptides fibres were sharper in the unwashed samples, indicating that the crystallinity or size of crystals decreased after the washing process. However, the XRD patterns of both the mineralised PCL/peptide samples after two weeks were very similar to those obtained after four weeks, which may indicate that there was no substantial increase in nucleation or growth of minerals after two weeks.

## 4. Conclusion

The electrospun PCL scaffolds supplemented with $P_{11}$-4 reported in this study revealed high cellular tolerability when L929 cells were seeded on the fibres, whereby an averaged cell viability of 84% and 100% was measured with respect to the DMEM and PCL control, respectively. It is thought that the incorporation of $P_{11}$-4 and $P_{11}$-8 into PCL scaffold nonwovens can effectively prevent their rapid dissolution in near-physiologic conditions. At least 40% of the $P_{11}$-8 peptide and 65% of the $P_{11}$-4 peptide remains within electrospun scaffolds after 7 days of incubation, even when pH conditions promote transformation of the self-assembled peptide into their respective monomeric state. At conditions close to biological pH, at least 75% of $P_{11}$-4 and 45% of $P_{11}$-8 peptides apparently remain in the scaffolds after 7 days of incubation. PCL electrospun scaffolds containing P11 peptides therefore have potential to modulate long-term therapeutic delivery to a bone defect site.

PCL/peptides scaffolds were also shown to promote apatite nucleation and subsequent growth of calcium phosphate crystals *in vitro*, compared with PCL only fibres. Apatite species nucleated on PCL/$P_{11}$-4 and PCL/$P_{11}$-8 scaffolds are characteristic of HAP after only one and four weeks respectively. It was found that $P_{11}$-4 loaded scaffolds showed better control over mineralisation even at decreased time periods, whilst HAP mineralisation was not observed on $P_{11}$-8 loaded scaffolds earlier than four weeks. These findings are in line with *in vivo* reports on the bone regeneration capability of both peptides, suggesting that both SAP-enriched fibre nonwovens could serve as regenerative scaffolds.[20] Following the promising *in-vitro* data reported herein, it is clear that *in-vivo* studies would be necessary to ascertain the feasibility of developing a fibrous device to fulfil the potential target applications proposed in this report.

## Conflicts of interest

There are no conflicts of interest to declare.

## Acknowledgments

The authors gratefully acknowledge financial support from the Alumni of the University of Leeds for the research scholarship awarded to RG and the support of the Clothworkers' Centre for Textile Materials Innovation for Healthcare. We are grateful to and wish to thank Dr. A. Aggeli for helpful discussions during the initial stages of this work.

# Graphical abstract

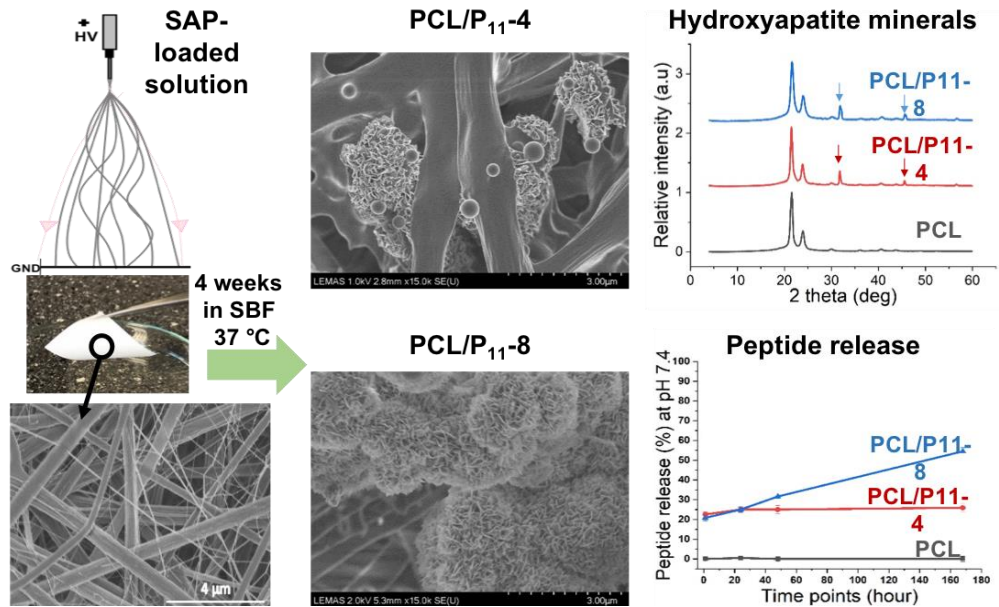

Structurally robust electrospun peptide-enriched scaffold, with controlled peptide release behaviour, support nucleation and growth of hydroxyapatite minerals *in vitro*.

**Table 1.** Sample nomenclature and formulation used in this study. All samples were prepared from a HFIP solution of PCL (6% w/w).

| SAMPLE ID | SAMPLE SPECIFICATION |
|:---:|:---:|
| 1 | 100% PCL |
| 2 | PCL/$P_{11}$-4 ([$P_{11}$-4]: 10 mg mL$^{-1}$) |
| 3 | PCL/$P_{11}$-4 ([$P_{11}$-4]: 20 mg mL$^{-1}$) |
| 4 | PCL/$P_{11}$-4 ([$P_{11}$-4]: 40 mg mL$^{-1}$) |
| 5 | PCL/$P_{11}$-8 ([$P_{11}$-8]: 40 mg mL$^{-1}$) |

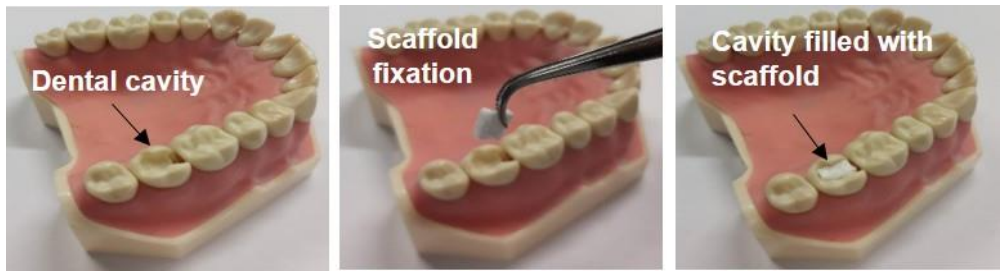

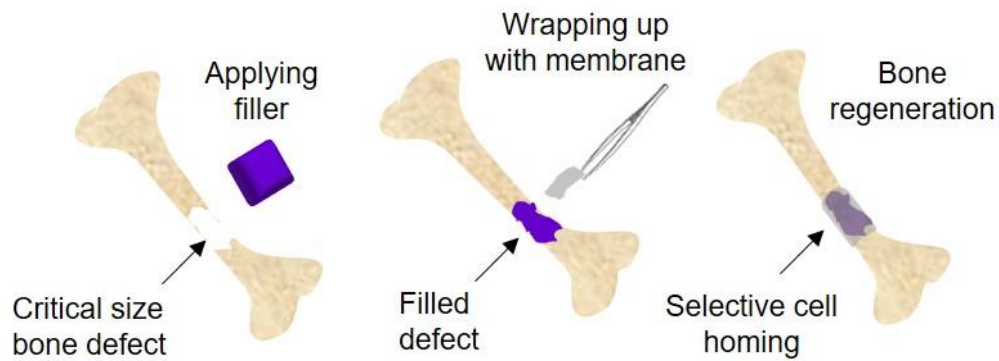

**Figure 1. Schematic of potential application of electrospun SAP loaded scaffolds; A) dental cavity management where the defect can be filed by a customised size and shape scaffold and it can then be fixed within the defect by e.g. using a biodegradable glue. B) Orthopaedic GBR membrane where the scaffold can be wrapped around an already filled defect to guide the bone tissue regeneration and prevent soft tissue intervention.**

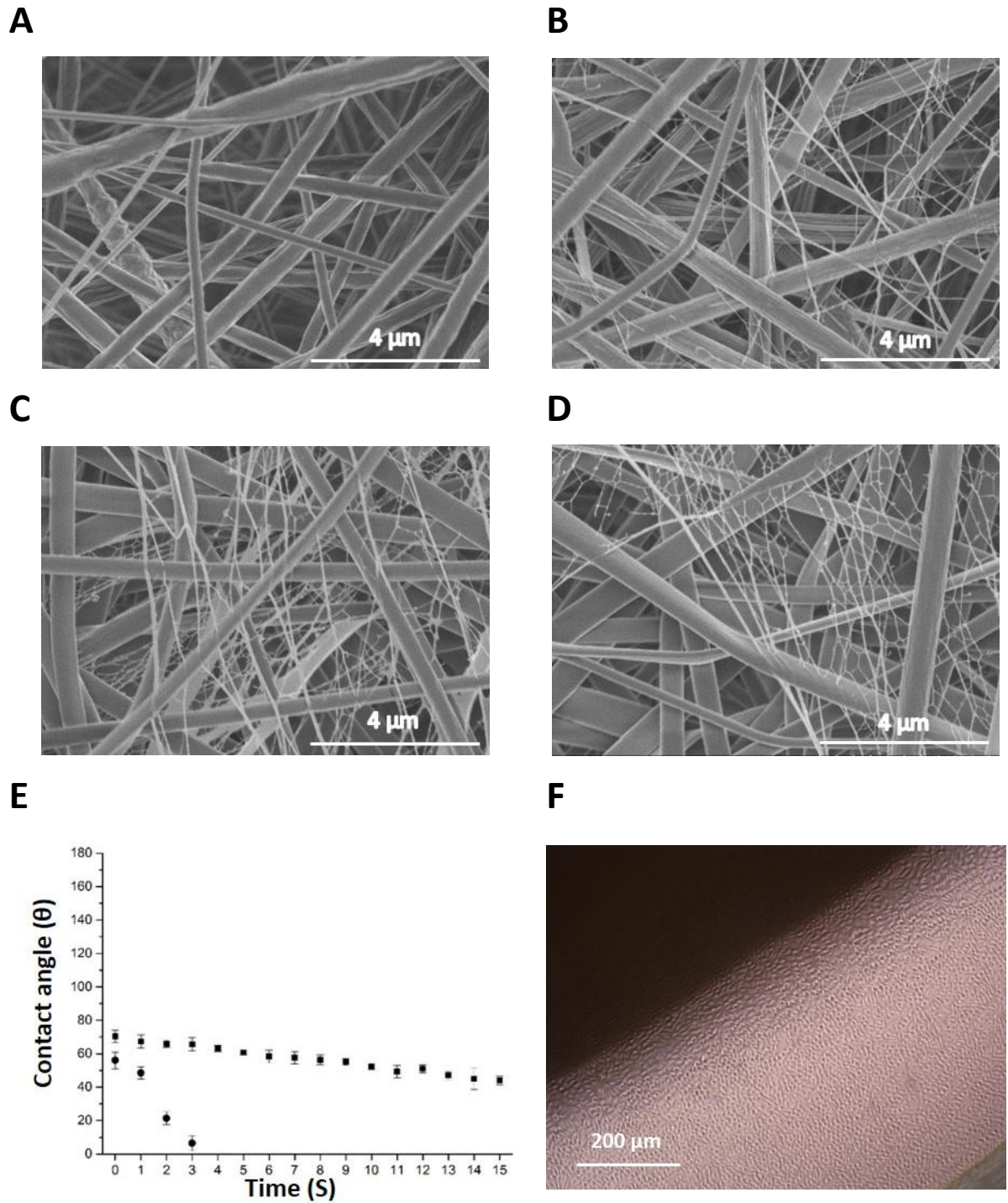

**Figure 2.** (A-D) SEM micrographs of PCL and $P_{11}$-4-supplemented nonwovens: (A) sample 1; PCL, (B-D) sample 2-4; PCL with $P_{11}$-4 concentration of 10, 20 and 40 mg mL$^{-1}$ respectively. (E) Dynamic water contact angle on electrospun scaffolds for 15 s of samples 2, and 3, PCL with $P_{11}$-4 concentration of 10 (■) and 20 mg mL$^{-1}$ (●) and (F) Light microscopy images of L929 cells in contact with sample 4, PCL with $P_{11}$-4 concentration of 40 mg mL$^{-1}$.

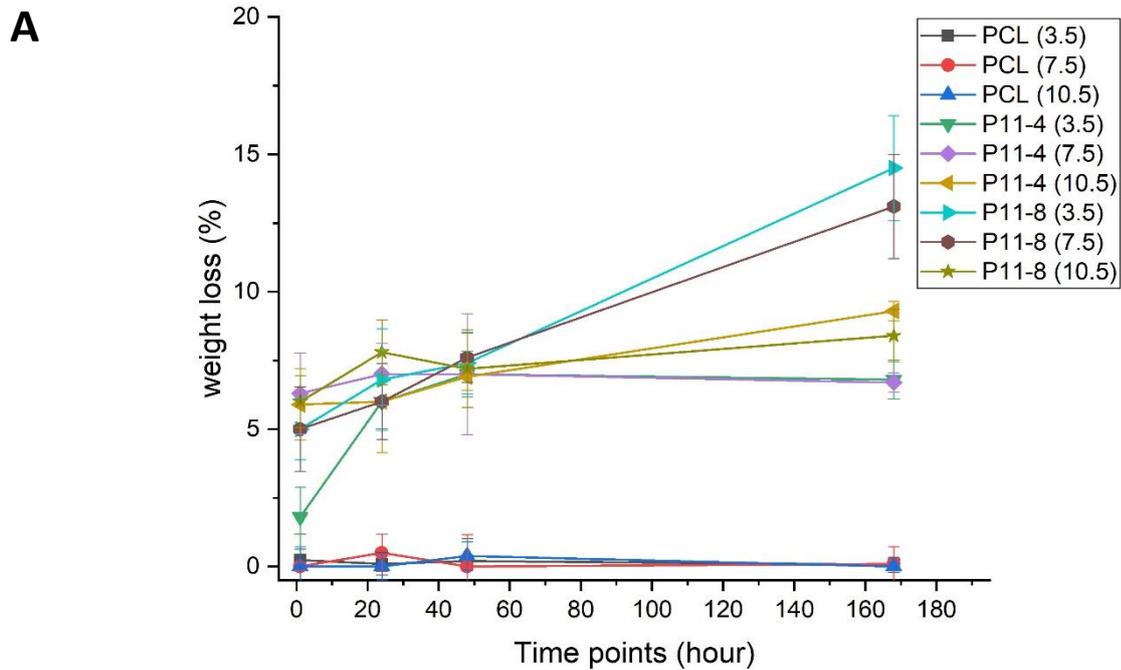

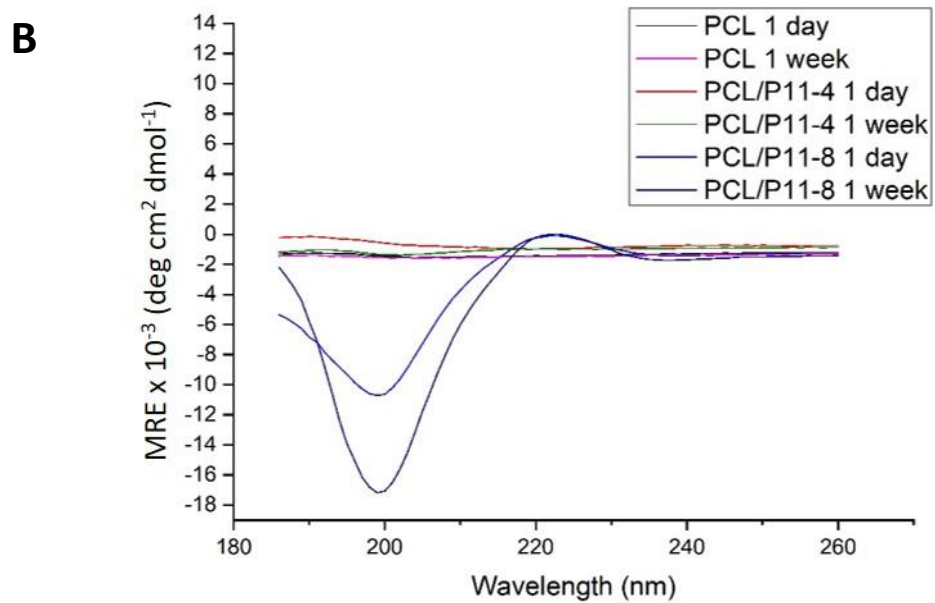

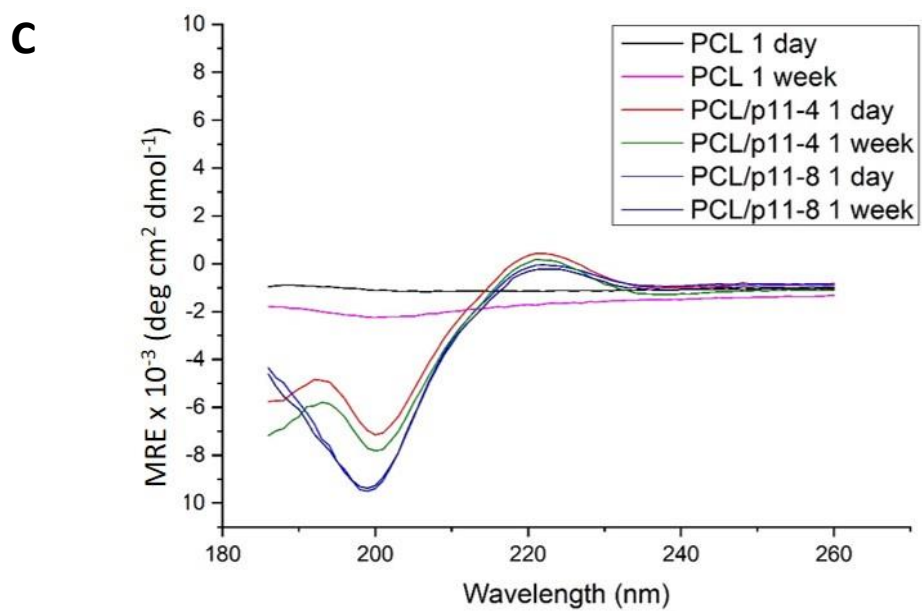

Figure 3. (A) Overall mass loss of peptide from sample 4 and 5 (PCL with $P_{11}$-4 and $P_{11}$-8 concentration of 40 mg mL$^{-1}$ respectively) in aqueous solution at different pH (A). (B-C) CD spectra of supernatant of fibrous samples 4 and 5 (PCL with $P_{11}$-4 and $P_{11}$-8 concentration of 40 mg mL$^{-1}$ respectively) dissolved in water for 1, 24, 48 and 168 hours at pH 3.5 and pH 10.5 respectively.

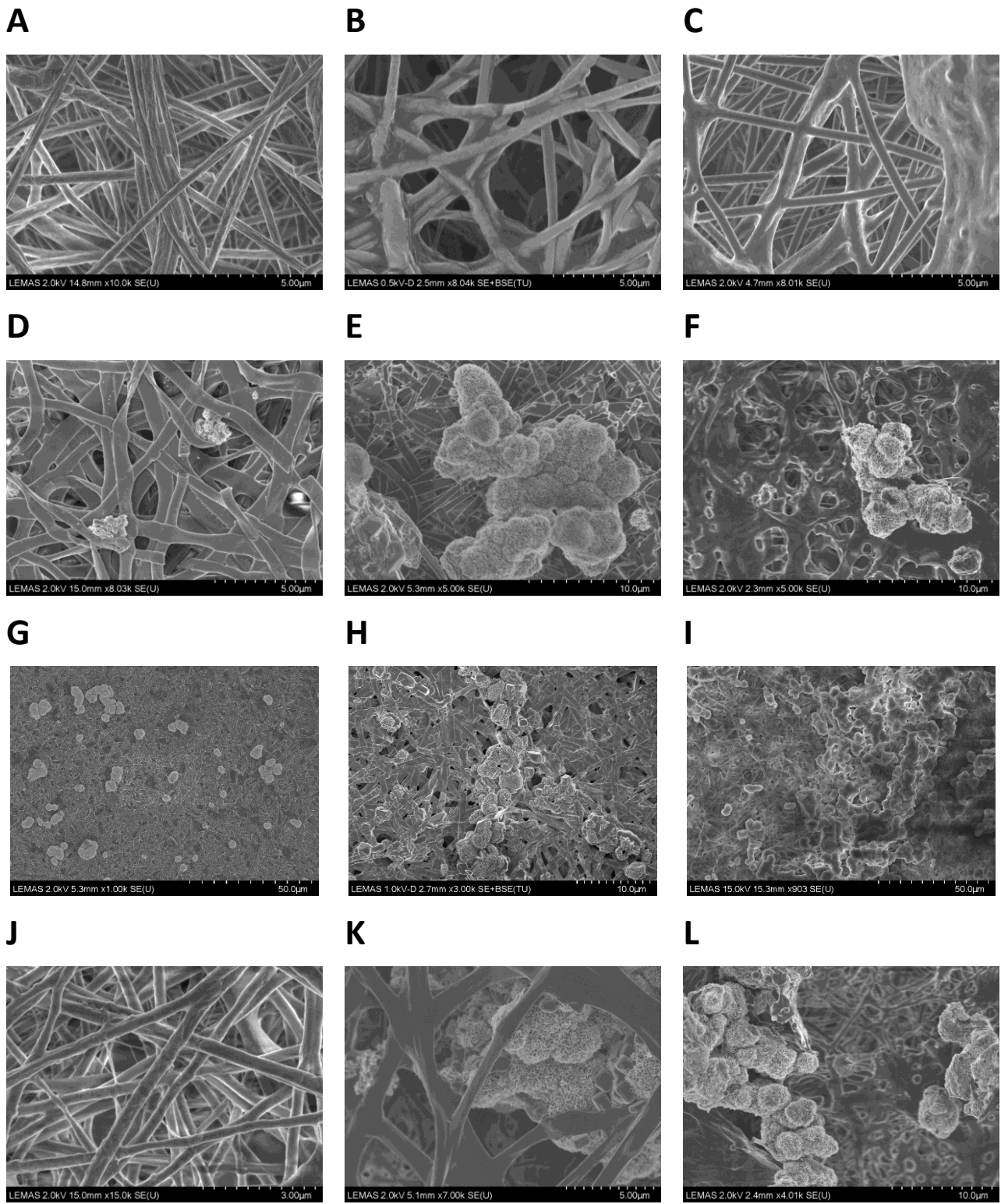

**Figure 4.** SEM micrographs of of electrospun scaffolds following incubation in SBF: (A-C) Sample 1; PCL after one , two and four weeks. (D-F) Sample 4, PCL with $P_{11}$-4 concentration of 40 mg mL$^{-1}$ after one week, two weeks and four weeks. (G-I) Low magnification SEM micrographs of sample 4, after one, two and four weeks incubation showing the distribution of crystals formed all over the fibres. (J-L) Sample 5, PCL with $P_{11}$-8 concentration of 40 mg mL$^{-1}$ after one , two and four weeks.

**Table 2. Mean Ca:P ratio of the electrospun scaffolds incubated in SBF for up to 4 weeks based on EDX analysis (concentration of $P_{11}$-4 and $P_{11}$-8 is the solution: 40 mg mL$^{-1}$).**

| Sample | Incubation time | | | | | |
|---|---|---|---|---|---|---|
| | Ca:P (1 week) | | Ca:P (2 weeks) | | Ca:P (4 weeks) | |
| | washed | unwashed | washed | unwashed | washed | unwashed |
| PCL | 0 | 0 | 0 | 0 | 0 | 0 |
| PCL supplemented with $P_{11}$-4 | 1.68 (SD=0.21) | 1.77 (SD=0.03) | 1.72 (SD=0.09) | 1.64 (SD=0.42) | 1.76 (SD=0.30) | 1.82 (SD=0.27) |
| PCL supplemented with $P_{11}$-8 | 0.01 (SD=1.3) | 0.31 (SD=1.8) | 0.50 (SD=0.7) | 1.50 (SD=0.1) | 0.21 (SD=0.33) | 1.65 (SD=0.39) |

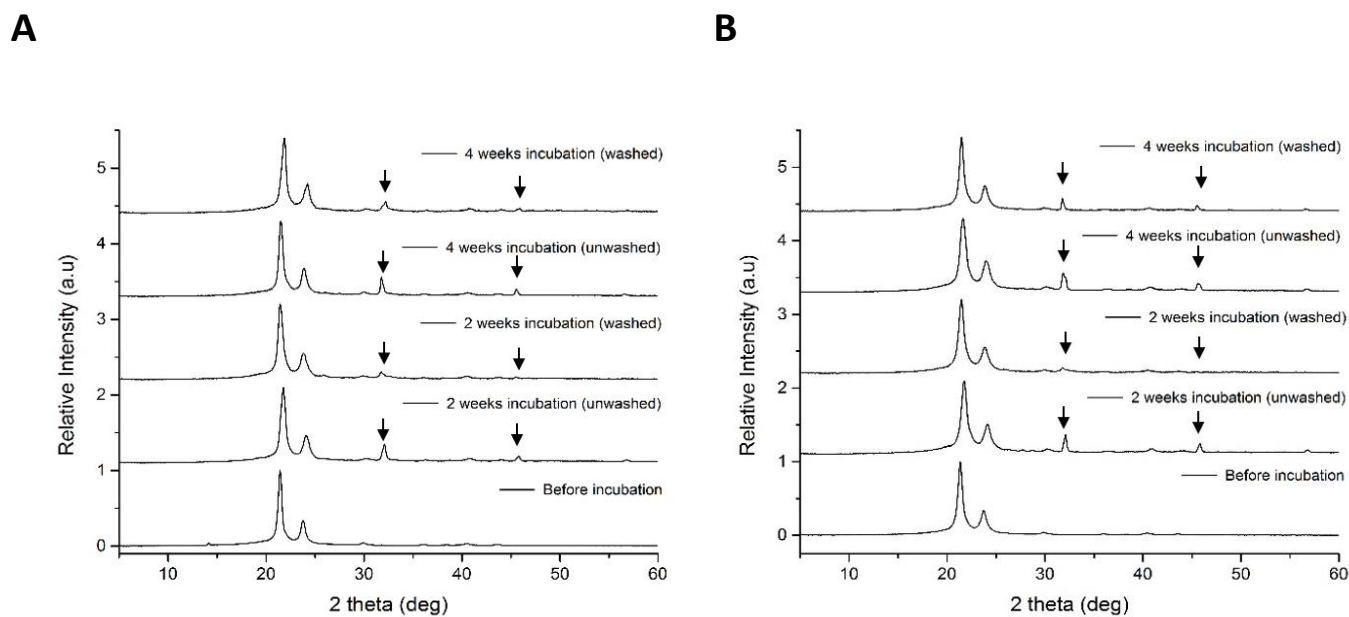

**Figure 5.** X-ray diffraction patterns of electrospun scaffolds before and after two and four weeks incubation in SBF (both washed and unwashed samples): (A) sample 4; PCL with $P_{11}$-4 concentration of 40 mg mL$^{-1}$ and (B) sample 5; PCL with $P_{11}$-8 concentration of 40 mg mL$^{-1}$.

# Supplementary information

**SI Table 1. Reagents for preparation of SBF in the required order of dissolution.**

| Order | Reagent | Amount | Purity (%) | Formula weight (g/mol) |
|---|---|---|---|---|
| 1 | Sodium chloride (NaCl) | 8.03 | 99.5 | 58.44 |
| 2 | Sodium hydrogen carbonate ($NaHCO_3$) | 0.35 | 99.5 | 84.00 |
| 3 | Potassium chloride (KCl) | 0.25 | 99.5 | 74.55 |
| 4 | Di-potassium hydrogen phosphate trihydrate ($K_2HPO_4 \cdot 3H_2O$) | 0.23 | 99 | 228.22 |
| 5 | Magnesium chloride hexahydrate ($MgCl_2 \cdot 6H_2O$) | 0.31 | 98 | 203.30 |
| 6 | Hydrochloric acid solution (HCl) = 1 mol/L | 39 (mL) | - | - |
| 7 | Calcium chloride ($CaCl_2$) | 0.29 | 95 | 110.98 |
| 8 | Sodium sulphate ($Na_2SO_4$) | 0.07 | 99 | 142.04 |
| 9 | Tris- hydroxymethyl aminomethane ($(HOCH_2)_3CNH_2$) | 6.11 | 99 | 121.13 |

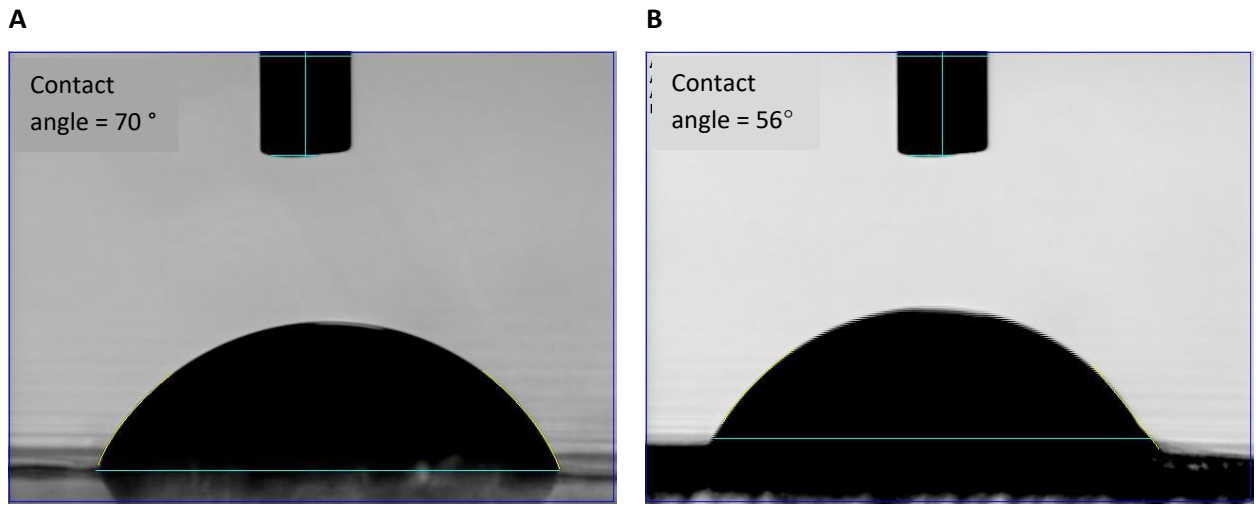

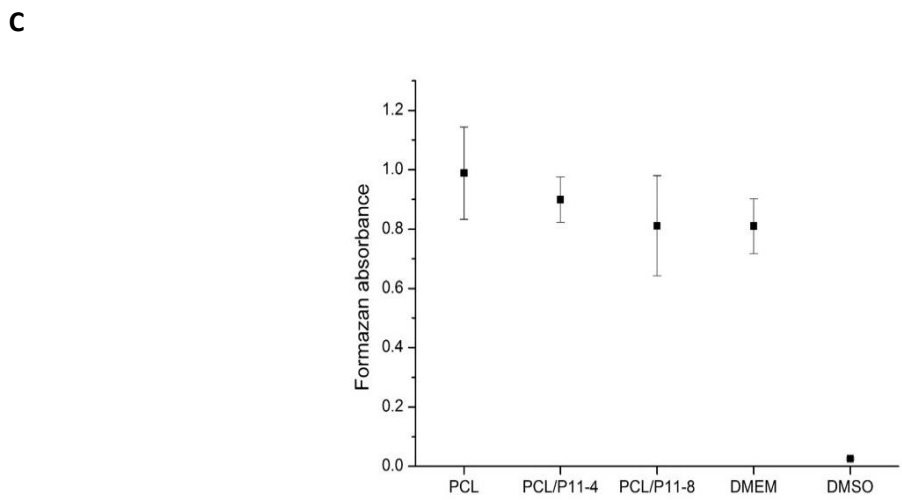

SI Figure 1. Initial contact angle of deionised water on electrospun scaffolds: (A-B) sample 2 and 3; PCL with 10 and 20 mg mL$^{-1}$ P$_{11}$-4 respectively, showing increase in hydrophilicity with the addition of peptide. (C) OD absorption of fibrous samples 4 and 5 (PCL with 40 mg mL$^{-1}$ of P$_{11}$-4 and P$_{11}$-8 respectively) and control samples at 570-650 nm correlated to the number of viable cells.

SI Table 2. Cell viability percentage of samples 4 and 5 (PCL with 40 mg mL$^{-1}$ of P$_{11}$-4 and P$_{11}$-8 respectively) calculated based on OD absorption data. Blank is the the cell-free DMEM sample.

| Sample | Mean OD (6 repetition) | Cell Viability % Based On Control 1 | Cell Viability % Based On Control 2 |
|---|---|---|---|
| **PCL/P11-4** | 1.21 | 100.12 | 83.67 |
| **PCL/P11-8** | 1.12 | 92.08 | 76.95 |
| **Blank** | 0.05 | - | - |
| **PCL (control 1)** | 1.21 | - | - |
| **Cell only (control 2)** | 1.43 | - | - |
| **Negative control** | 0.10 | - | - |

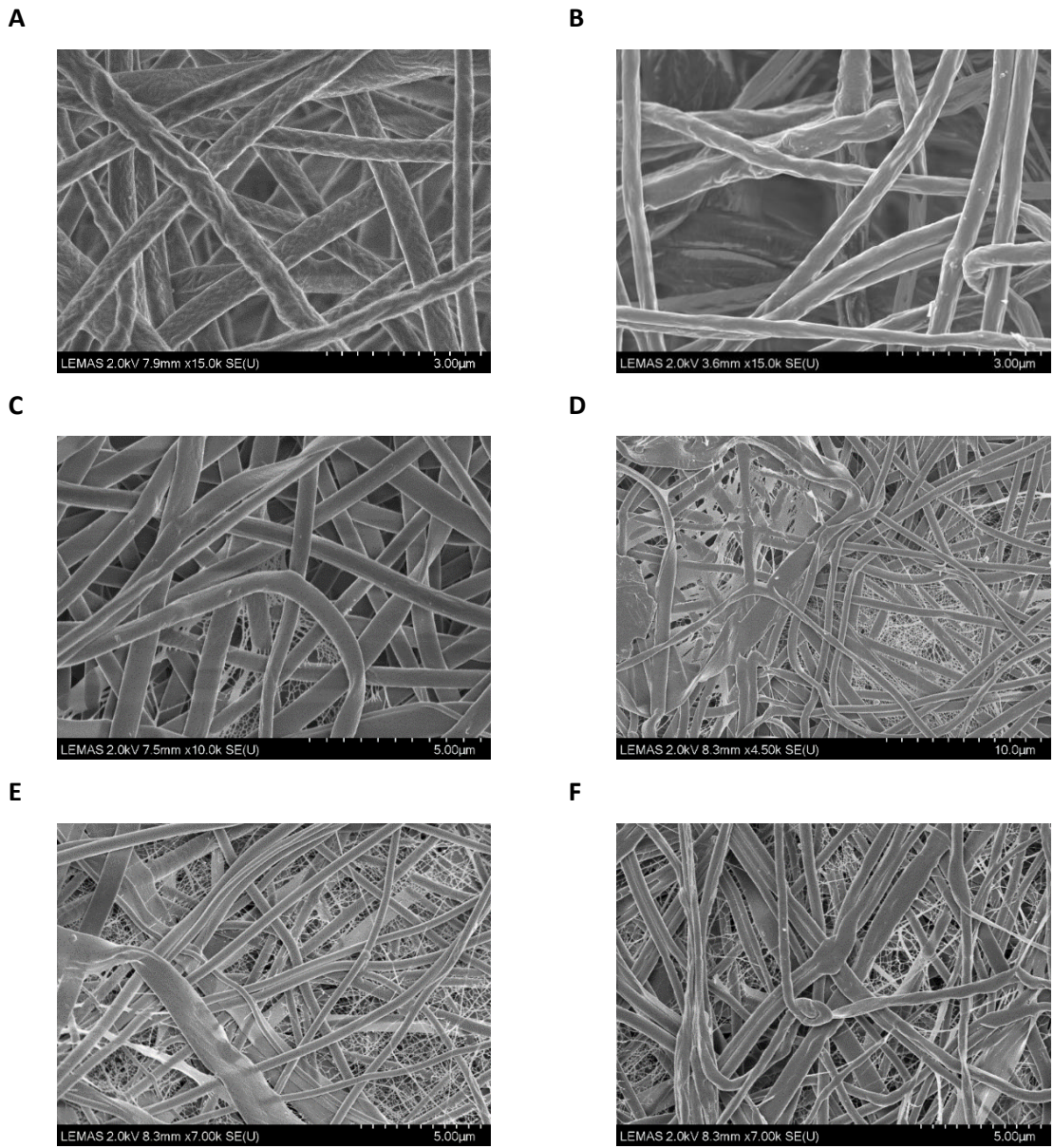

**SI Figure 2.** SEM micrographs of electrospun scaffolds after 168 hours incubation of (A-B) PCL at pH 3.5 and 10.5, (C-D) PCL/$P_{11}$-4 (40 mg mL$^{-1}$) at pH 3.5 and 10.5, and (E-F) PCL/$P_{11}$-8 (40 mg mL$^{-1}$) at pH 3.5 and 10.5.

**SI Table 3.** An example of elemental analysis of chemical elements in the P$_{11}$-4-supplemented scaffolds (with peptide concentration of 40 mg mL$^{-1}$) after immersion in SBF for 2 weeks obtained by SEM/EDX showing calcium and phosphorous atomic contents.

| Element | Atomic Content (%) |
|---|---|
| C | 37.07 |
| O | 40.54 |
| Na | 1.40 |
| Mg | 0.44 |
| P | 6.85 |
| Cl | 1.61 |
| K | 0.09 |
| Ca | 11.80 |
| Cu | 0.02 |
| Ir | 0.19 |
| Total: | 100.00 |

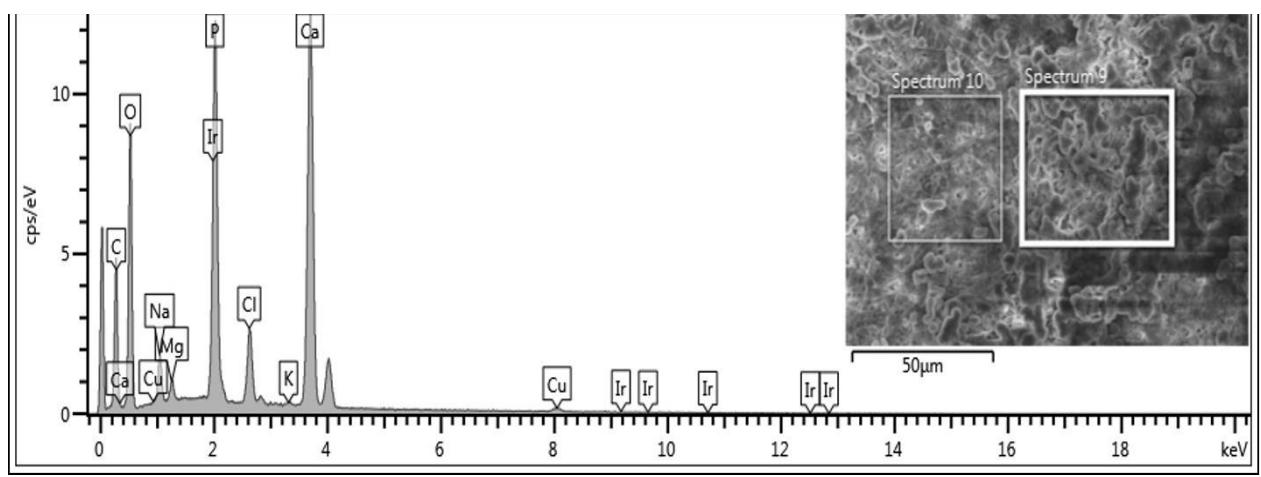

**SI Figure 3.** EDX spectra of P$_{11}$-4-supplemented scaffold (with peptide concentration of 40 mg mL$^{-1}$) immersed in SBF for 2 weeks showing calcium and phosphorous peaks.

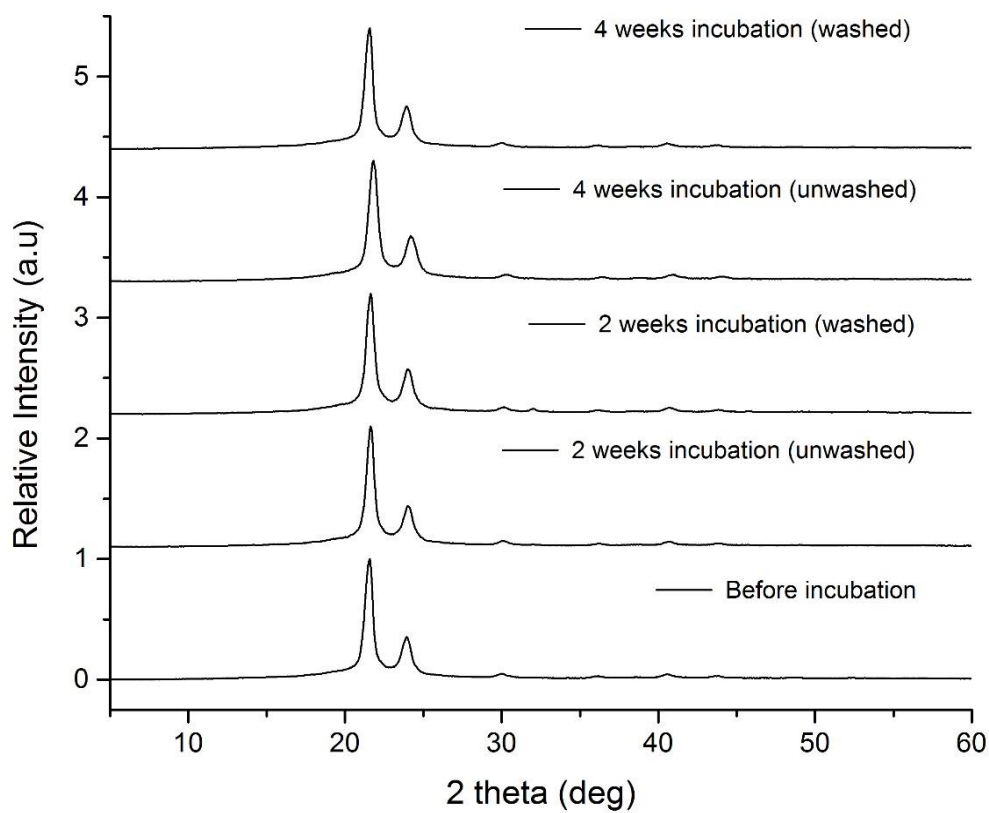

**SI Figure 4.** X-ray diffraction patterns of PCL control scaffolds before and after 2 and 4 weeks incubation in SBF (both washed and unwashed samples) showing no hydroxyapatite formation.